\documentclass{optica-article}

\journal{opticajournal} 

\articletype{Research Article}

\usepackage{lineno}

\begin{document}

\title{1.56 THz Full-Field Rydberg-Atom Imaging with 0.3 mm-scale Spatial Response}

\author{
Xianzhe Li,\authormark{1,2}
Yibin Huang,\authormark{4}
Zhiwen Chen,\authormark{2,3}
Kaiqing Zhang,\authormark{1,*}
and Haixiao Deng\authormark{1,*}
}

\address{
\authormark{1}Shanghai Advanced Research Institute, Chinese Academy of Sciences, Shanghai 201210, China\\
\authormark{2}University of Chinese Academy of Sciences, Beijing 100049, China\\
\authormark{3}Shanghai Institute of Applied Physics, Chinese Academy of Sciences, Shanghai 201800, China\\
\authormark{4}South China Normal University, Guangzhou 510631, China
}

\email{\authormark{*}zhangkq@sari.ac.cn}
\email{\authormark{*}denghx@sari.ac.cn}


\begin{abstract*} 
THz imaging via Rydberg-atom up-conversion offers a powerful full-field modality with high sensitivity and rapid acquisition. However sub-millimeter imaging above 1 THz is still constrained by source power and beam-delivery loss. In this paper, we demonstrate a \(50~\mathrm{mm} \times 50~\mathrm{mm}\) full-field imaging at 1.56~THz by implementing the \(11P_{3/2} \rightarrow 10D_{5/2}\) transition of a cesium Rydberg-atom vapor. The measured spatial response FWHMs are \(0.314~\mathrm{mm}\) in horizontal and \(0.354~\mathrm{mm}\) in vertical for a \(0.5~\mathrm{mm}\) pinhole, alongside a sensitivity of \(31.7~\mathrm{fW}\,\mathrm{s}^{-1/2}/(6.5 \times 6.5)~\mu\mathrm{m}^{2}\), and a frame rate up to \(300~\mathrm{fps}\). Under matched illumination, relay, and image-processing conditions, mask imaging at 1.56 THz reproduces finer features and edges with substantially higher edge contrast, deeper intensity-trough separation in line profiles, and recovered detail that remains blurred or unresolved at 0.55 THz. Furthermore, preliminary imaging of a \(200~\mu\mathrm{m}\) tissue slice suggests morphological correspondence with H\&E histology histology, offered as a proof-of-concept for label-free tissue characterization. By validating the wavelength-dependent resolution scaling in Rydberg THz imaging, this work establishes frequency scaling as a general design principle for sub-millimeter Rydberg-THz imaging.

\end{abstract*}

\section{INTRODUCTION}

THz wave sits between the electronic and optical domains and can penetrate plastics, paper, textiles, and many dielectric or biological materials without the ionization risk of X-rays, which makes it attractive for nondestructive testing, security screening, and biomedical inspection\cite{bib1,bib2,bib3,bib4,bib5}. In practice, however, THz imaging is hard to combine with full-field operation, high frame rate, and fine spatial detail at the same time. The difficulty becomes sharper as the operating frequency is raised because the diffraction-limited spot size improves roughly linearly with frequency, but available continuous-wave THz source power falls much faster, so the nominal resolution gain is rarely recovered in a real imaging system. Any route to sub-millimeter full-field THz imaging therefore has to solve not only detection sensitivity but also field delivery and power budget.



Building upon advances in Rydberg-state optical detection, real-time near-field THz imaging using atomic optical fluorescence was demonstrated, establishing a powerful and promising approach for converting THz fields into detectable optical images \cite{bib6,bib7,bib8,bib9,bib10,bib11,bib12}. This approach was further developed through full-field THz imaging using cesium vapor at 0.55 THz, achieving near-diffraction-limited spatial response over a roughly \(10~\mathrm{mm} \times 10~\mathrm{mm}\) sensing area and video-rate acquisition up to 3000 fps \cite{bib13}. Following works at 0.55 THz, including large-field \(50~\mathrm{mm} \times 50~\mathrm{mm}\) imaging \cite{bib14}, dual-camera high-speed readout \cite{bib15,bib16}, and AI-assisted resolution enhancement \cite{bib17}, pushed field size, frame rate, and effective sensitivity forward. Reflection-mode geometries for bulk-surface inspection were also explored as an extension \cite{bib18}. Nevertheless, the full-field spatial response in these systems remained at the millimeter scale. More recent Rydberg demonstrations operating above 1 THz have shown real-time detection and improved sensitivity, but still report an imaging response of about 1.7 mm \cite{bib19,bib20}. The sub-millimeter full-field regime above 1 THz has therefore remained unaddressed because the bottleneck is not the absence of atomic detection, but the fact that higher frequency alone loses signal to source power and propagation loss before the smaller diffraction scale can be used.

In this work, we demonstrate a high-resolution cesium Rydberg-atom-based THz imaging system by exploiting the \(11P_{3/2} \rightarrow 10D_{5/2}\) transition at 1.56 THz. By systematically optimizing the candidate Rydberg transitions by the figure of merit and thus balance the diffraction-limited resolution, optical excitation efficiency, THz coupling strength, and fluorescence yield, we establish a practical pathway toward high-resolution THz imaging. To preserve the THz field before upconversion, the beam is delivered through a low-loss TPX Fresnel optics path with matched illumination geometry, rather than relying on free-space propagation at higher frequency. The result shows that sub-millimeter spatial response in atomic THz imaging becomes accessible when higher frequency is combined with low-loss delivery and matched illumination, establishing frequency scaling as a design principle for submillimeter Rydberg-THz imaging.

\section{METHODS}

In a Rydberg-atom THz imaging system, the THz field first illuminates the imaging sample, thereby encoding its spatial information. The transmitted field is then guided and focused onto the atomic vapor by the THz optical system, where the THz field is converted into readily detectable visible fluorescence. While this conversion enables sensitive, full-field optical readout, it cannot recover spatial frequencies that have already been lost during THz propagation. The fluorescence image therefore records the THz intensity distribution at the atomic interaction plane, and the fundamental far-field resolution remains governed by the THz wavelength and the numerical aperture (NA) of the THz optics.

For a diffraction-limited circular aperture with $\mathrm{NA}=n\sin\theta$, the intensity point-spread function (PSF) in the atomic plane is \cite{bib21}
\begin{equation}
I(r)=I_0\left[\frac{2J_1(k_{\mathrm{THz}}\mathrm{NA}\,r)}{k_{\mathrm{THz}}\mathrm{NA}\,r}\right]^2,
\label{eq:airy_psf}
\end{equation}
where $J_1$ is the first-order Bessel function, $r$ is the radial coordinate, and $k_{\mathrm{THz}}=2\pi/\lambda_{\mathrm{THz}}$ is the THz wave number. The first zero of $J_1(x)$ occurs at $x=3.8317$. Hence, the radius of the central Airy disk is
\begin{equation}
r_0=\frac{3.8317}{k_{\mathrm{THz}}\mathrm{NA}}\simeq0.61\frac{\lambda_{\mathrm{THz}}}{\mathrm{NA}}.
\label{eq:airy_radius}
\end{equation}
According to the Rayleigh criterion, the minimum resolvable separation between two point objects is therefore
\begin{equation}
\delta_{\mathrm{diff}}\simeq0.61\frac{\lambda_{\mathrm{THz}}}{\mathrm{NA}}
=0.61\frac{c}{f_{\mathrm{THz}}\mathrm{NA}}.
\label{eq:rayleigh}
\end{equation}
Equation~\eqref{eq:rayleigh} makes the role of operating frequency explicit. For a fixed optical geometry, increasing $f_{\mathrm{THz}}$ shortens the wavelength and proportionally decreases the diffraction-limited feature size. Importantly, this improvement occurs prior to THz-to-optical conversion: the atomic fluorescence serves as the readout channel and cannot recover spatial information lost through diffraction. Increasing the THz frequency is therefore a direct physical route to improved far-field resolution.

To realize a higher-frequency Rydberg-atom THz imaging system, a suitable imaging transition must satisfy several critical requirements: efficient optical preparation of the Rydberg state, strong THz coupling, efficient conversion of the THz-induced population redistribution into visible fluorescence, and sufficiently powerful THz source \cite{bib20,bib21,bib22}. Consequently, we systematically screened candidate Cs transitions within the accessible 1-3 THz range based on quantum-number and angular-momentum selection rules, laser accessibility, measured THz-source bands, transition dipole moments, and radiative decay rates. 

\begin{figure}[htbp]
    \centering
    \includegraphics[width=15cm]{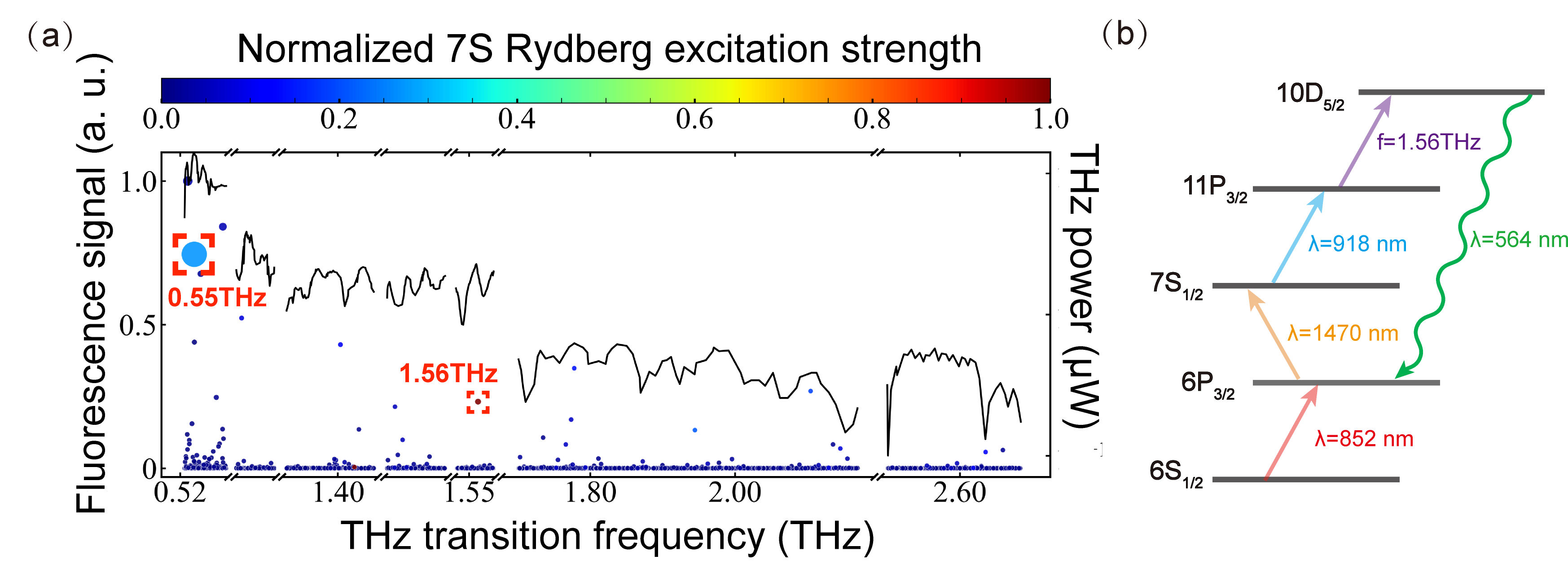}
    \caption{(a) Comprehensive performance distribution of candidate THz transitions. The blue points represent transitions satisfying the quantum-number constraints, angular-momentum selection rules, and available source-frequency range. The horizontal axis gives the transition frequency, and the vertical axis gives the normalized signal metric $S$.Marker color represents the normalized 7S-to-Rydberg excitation strength, whereas marker area scales with the relative FOM. The representative 0.55 THz ($14P_{3/2}\rightarrow13D_{5/2}$) and 1.56 THz ($11P_{3/2}\rightarrow10D_{5/2}$) transitions are highlighted. The black curves show the interpolated THz-source power on a logarithmic scale; gaps indicate discontinuous source bands. All visual encodings are normalized and are intended only for relative comparison. (b) Five-level excitation scheme for 1.56 THz detection. 
Three optical fields at 852, 1470, and 918~nm sequentially excite atoms from $6S_{1/2}$ to the Rydberg state $11P_{3/2}$. The 1.56~THz field couples $11P_{3/2}$ to $10D_{5/2}$, resulting in fluorescence emission at 564~nm.}
\end{figure}

For each candidate transition, the fluorescence response produced by the THz field is characterized by \cite{bib21}
\begin{equation}
S=d_{\mathrm{THz}}^2\left|F_{\mathrm{on}}-F_{\mathrm{off}}\right|,
\label{eq:signal_metric}
\end{equation}
where $d_{\mathrm{THz}}$ is the THz transition dipole moment, and $F_{\mathrm{on}}$ and $F_{\mathrm{off}}$ are the calculated visible-photon yields with the THz field on and off, respectively. To compare transition with realistic THz source, we define the overall figure of merit as
\begin{equation}
\mathrm{FOM}=P_{\mathrm{THz}}d_{\mathrm{Ryd}}^2S,
\label{eq:fom}
\end{equation}
where $P_{\mathrm{THz}}$ is the available source power at the candidate-transition frequency and $d_{\mathrm{Ryd}}$ is the dipole moment for the $7S_{1/2}$-to-Rydberg excitation step. The FOM therefore combines the three factors that directly determine the useful imaging signal: optical preparation, THz-driven population transfer, and available THz power. A high frequency improves the diffraction limit, but it is valid only when these signal-generation factors remain sufficiently strong.

Fig.~1(a) maps the comprehensive performance landscape of candidate transitions. The abscissa and ordinate represent the transition frequency and the normalized signal metric $S$, respectively. Marker color and area encode the normalized ground-to-Rydberg excitation strength and the relative FOM, with the interpolated THz-source power superimposed as a black curve. Because these metrics are normalized, this multi-parameter mapping serves to benchmark the relative strengths of candidate transitions rather than reflect absolute signal amplitudes.

Guided by this multi-parameter framework, the excitation ladder $6S_{1/2}\rightarrow6P_{3/2}\rightarrow7S_{1/2}\rightarrow11P_{3/2}$ depicted in Fig.~1(b) emerges as the optimal pathway. The large reduced matrix element for the final optical step, $\langle7S_{1/2}\Vert er\Vert11P_{3/2}\rangle=0.270\,ea_0$, ensures highly efficient population transfer to the Rydberg state, given that the excitation rate scales approximately as $W_3\propto d_3^2P_3$. The subsequent THz transition $11P_{3/2}\rightarrow10D_{5/2}$ operates at 1.562~THz. It further features a $10D_{5/2}$ state that exhibits a pronounced direct radiative decay rate to $6P_{3/2}$ of $\Gamma_{10D_{5/2}\rightarrow6P_{3/2}}\approx1.81\times10^6\,\mathrm{s}^{-1}$, thereby facilitating rapid and efficient conversion of the THz-induced population redistribution into visible fluorescence. Notably, the requisite excitation wavelength of approximately 918.7 nm is readily accessible using tapered-amplifier system.

As highlighted in Fig.~1(a), rate-equation modeling underscores the distinct advantages of the 1.56 THz scheme over the conventional 0.55 THz transition \cite{bib13,bib15}. Within our evaluation framework, the optical excitation dipole moment increases from approximately $0.0574\,ea_0$ at 0.55~THz to $0.1102\,ea_0$ at 1.56~THz, accompanied by an enhanced calculated fluorescence-yield variation from
$0.1835$ to $0.2256$. 
Although a higher-frequency transition near
$2.46\,\mathrm{THz}$ would theoretically decrease
the diffraction-limited feature size, the available frequency-multiplier
chain supplies only approximately $0.1\,\mu\mathrm{W}$,
severely constraining the usable THz field strength, corresponding to a FOM more than one order of magnitude below the 1.56 THz channel. Consequently, the
$1.56\,\mathrm{THz}$ transition optimizes the trade-offs among
diffraction-limited resolution, optical preparation efficiency,
fluorescence-readout yield, and available THz drive power.

The experimental setup is designed to achieve full-field imaging of a 1.56 THz field via Rydberg-atom-mediated THz-to-visible conversion within a cesium vapor cell. A three-step optical excitation scheme is employed to prepare the cesium atoms in the $11P_{3/2}$ Rydberg state. Specifically, an 852 nm probe laser drives the $6S_{1/2}(F=4)\rightarrow6P_{3/2}(F=5)$ transition, followed by a 1470 nm coupling laser addressing the $6P_{3/2}(F=5)\rightarrow7S_{1/2}(F=4)$ transition. A final Rydberg laser at approximately $918~\mathrm{nm}$ then promotes the atoms from $7S_{1/2}$ to $11P_{3/2}$. Following spatial shaping and alignment, the three laser beams are combined and spatially overlapped within the vapor cell to establish a thin, two-dimensional excitation region that serves as the active sensing plane for THz imaging. A schematic of the experimental setup is shown in Fig. 2(a). The THz beam profile at the imaging plane is presented in Fig. 2(b).

\begin{figure}[htbp]
\centering\includegraphics[width=12cm]{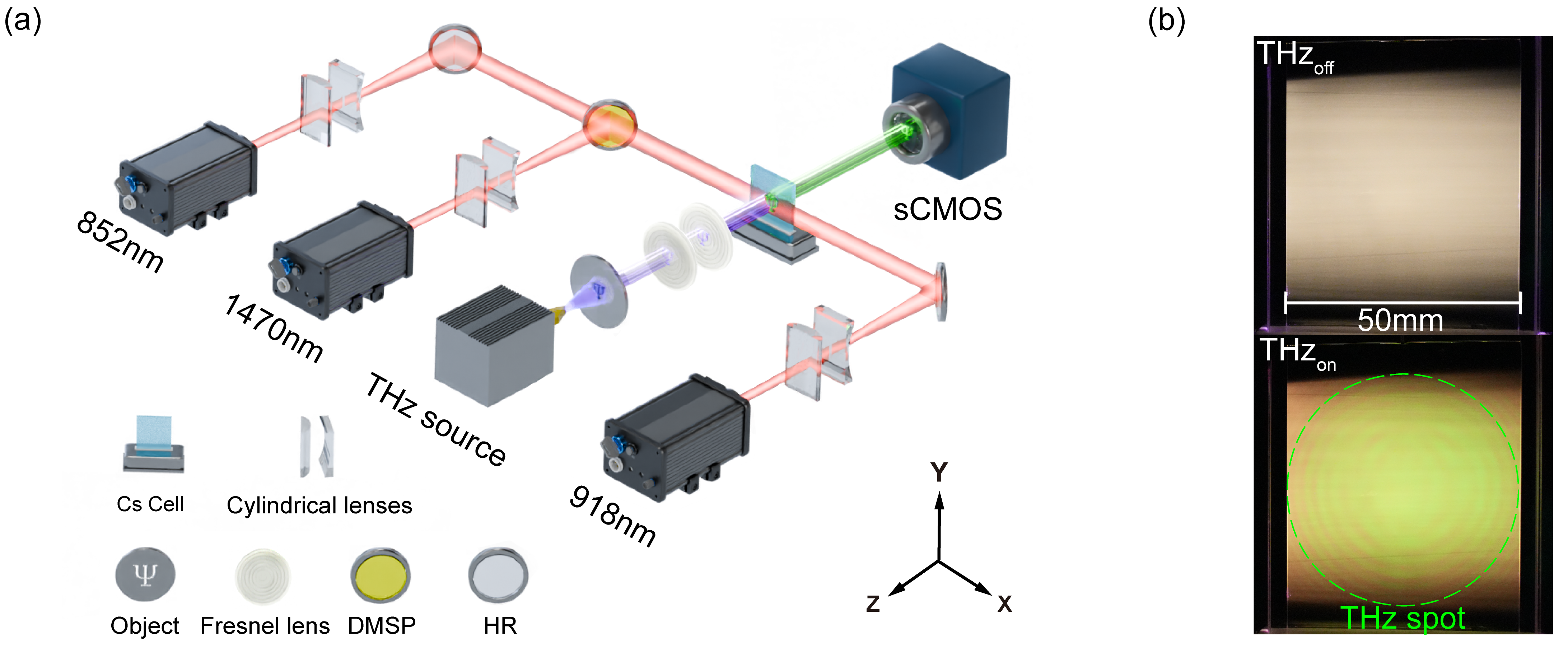}
\caption{Experimental configuration. (a) The three excitation beams are spatially overlapped in the Cs vapor cell, and the THz beam is focused onto the atomic interaction region. The resulting 564 nm fluorescence is collected from the side and imaged onto a camera. (b) THz-induced fluorescence response of the atomic vapor. The THz-induced fluorescence enhancement reveals the beam profile over a $50 \times 50$ mm$^{2}$ imaging area, with the THz spot indicated by the green dashed circle.}
\end{figure}

Because available THz power drops rapidly at frequencies above 1 THz, mitigating propagation losses is critical. Consequently, the beam-delivery optics were specifically optimized for 1.56 THz. Unlike conventional bulk Teflon lenses used in lower-frequency systems, we employ a TPX Fresnel lens to shape the incident radiation. TPX exhibits exceptional optical transparency and minimal dielectric loss in the THz domain, while the Fresnel configuration substantially reduces the bulk dielectric volume. This design minimizes internal absorption and shortens the effective propagation path through the lens material. Together, the intrinsically low-loss TPX and minimized footprint effectively mitigate attenuation, ensuring that the limited source power is maximized at the atomic sensing region for robust fluorescence imaging.

The THz beam is collimated to illuminate the sample, after which the object's transmitted field is projected onto the Rydberg excitation plane within the vapor cell via a custom THz imaging lens. To exploit the high-resolution advantages analyzed above, a radiation frequency of 1.562 THz is applied, driving the resonant $11P_{3/2}\rightarrow10D_{5/2}$ transition. This spatial field distribution is thus mapped into a proportional Rydberg population distribution. Subsequent radiative cascade produces visible fluorescence, which is collected orthogonally to the THz axis and recorded by a highly sensitive scientific camera (Dhyana 400BSI V3). An optical band-pass filter is positioned before the camera to suppress scattered excitation light and background radiation, thereby significantly enhancing the fluorescence signal-to-noise ratio. Ultimately, the complete two-dimensional THz field distribution is captured in a single exposure without requiring mechanical raster scanning.

\section{RESULTS}

\begin{figure}[htbp]
\centering\includegraphics[width=12cm]{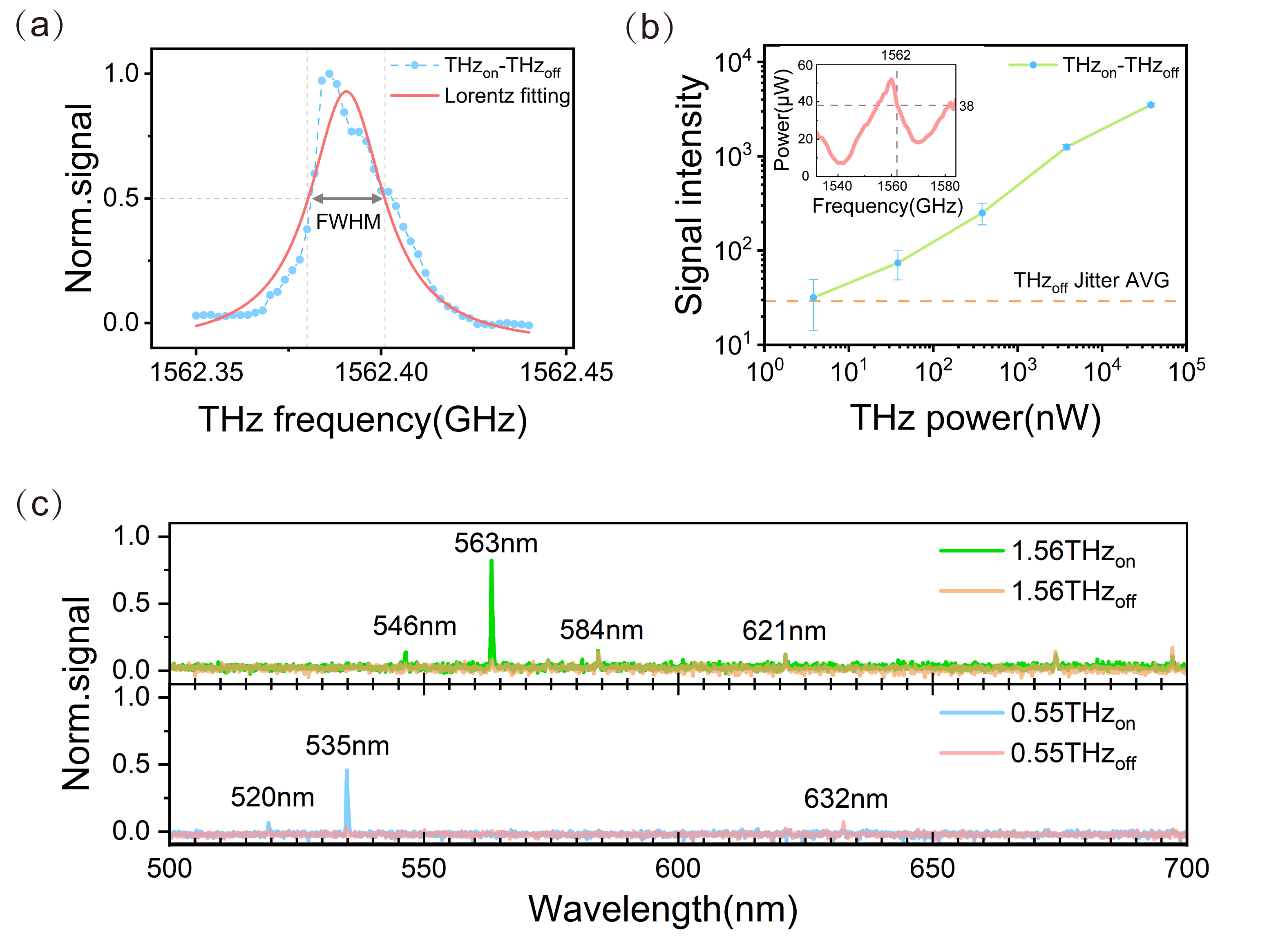}
\caption{Fluorescence characterization of the 1.56 THz imaging system. (a) Normalized differential fluorescence signal ($\mathrm{THz}_{\rm on}-\mathrm{THz}_{\rm off}$) as a function of THz frequency. The solid red curve represents a Lorentzian fit to the measured data. (b) Differential fluorescence signal as a function of incident THz power. The inset shows the measured output power spectrum of the THz source, with an output power of approximately 38~$\mu$W. (c) Normalized fluorescence spectra of the atomic vapor with ($\mathrm{THz}_{\rm on}$) and without ($\mathrm{THz}_{\rm off}$) THz excitation at 1.56 THz (upper panel) and 0.55 THz (lower panel).}
\end{figure}

To characterize the frequency response and detection sensitivity of the 1.56 THz imaging system, the THz frequency was tuned across the Rydberg transition while monitoring the fluorescence intensity, defined as $\mathrm{THz}_{\rm on}-\mathrm{THz}_{\rm off}$. As shown in Fig.~3(a), the fluorescence signal peaks near resonance and drops sharply as the THz frequency detunes. Fitting the measured response yields a FWHM of $24.37~\mathrm{MHz}$, which defines the effective detection bandwidth of the imaging system.

The sensitivity and minimum detectable power (MDP) of the imaging system were evaluated by progressively attenuating the incident THz power, with the results presented in Fig.~3(b). The inset depicts the measured output power spectrum of the THz source near $1.56~\mathrm{THz}$, showing an output power of approximately $38~\mu\mathrm{W}$. Using calibrated attenuators, the incident THz power was systematically reduced in 10-fold steps while recording the corresponding normalized differential fluorescence signal ($\mathrm{THz}{\mathrm{on}}-\mathrm{THz}{\mathrm{off}}$). The background noise floor was established from fluctuations measured under the $\mathrm{THz}_{\mathrm{off}}$ condition. The signal remained clearly distinguishable down to an incident THz power of $38~\mathrm{nW}$, where the THz beam forms an approximately circular illumination zone with a diameter of $20~\mathrm{mm}$. For a $2048\times2048$-pixel camera, this region spans approximately $5.27\times10^{5}$ pixels. Assuming a uniform intensity distribution across the illuminated area, the MDP is determined to be approximately $104~\mathrm{fW/pixel}$ at an exposure time of $200~\mathrm{ms}$. Scaling inversely with the square root of the integration time, the resulting time-normalized MDP is estimated to be approximately $31.7~\mathrm{fW,s^{-1/2}}$ per $(6.5\times6.5)~\mu\mathrm{m}^{2}$, demonstrating the sensitivity of the $1.56$-THz Rydberg imaging system.

\begin{figure}[htbp]
\centering\includegraphics[width=12cm]{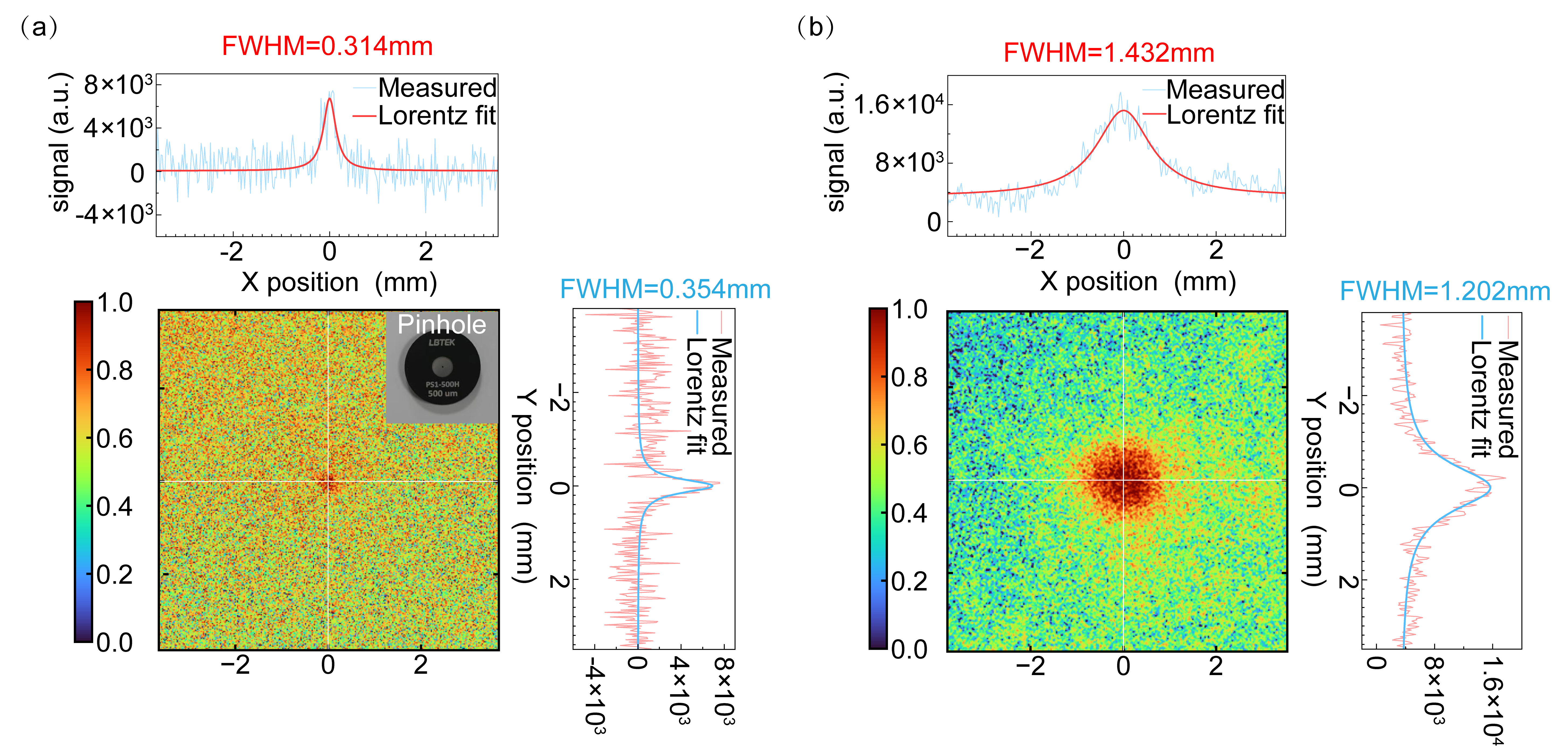}
\caption{Experimental characterization of the spatial response of the
    THz imaging system using a \(0.5~\mathrm{mm}\)-diameter pinhole. The
    upper and lower rows show the imaging results obtained at
    \(1.56~\mathrm{THz}\) and \(0.55~\mathrm{THz}\), respectively. The two-dimensional images indicate the
    positions at which the horizontal and vertical intensity profiles were
    extracted. }
\end{figure}

To characterize the spectral properties of the fluorescence, emission spectra under the $\mathrm{THz}{\mathrm{on}}$ and $\mathrm{THz}{\mathrm{off}}$ conditions were recorded using a spectrometer, as displayed in Fig.~3(c). Because the $1.56$-THz and $0.55$-THz fields couple with distinct Rydberg transitions, they generate markedly different fluorescence profiles. Specifically, the $1.56$-THz excitation (upper panel) produces prominent emission peaks at $546~\mathrm{nm}$, $563~\mathrm{nm}$, $584~\mathrm{nm}$, and $621~\mathrm{nm}$, whereas the conventional $0.55$-THz scheme (lower panel) yields enhanced peaks at $520~\mathrm{nm}$, $535~\mathrm{nm}$, and $632~\mathrm{nm}$. A magnified view further reveals that the fluorescence intensity under the $\mathrm{THz}_{\mathrm{on}}$ condition peaks near $564~\mathrm{nm}$, which is in excellent agreement with the theoretically predicted wavelength.

To quantitatively evaluate the spatial resolution of the imaging system, a metallic mask featuring a circular pinhole with a diameter of 0.5 mm  was placed at the object plane and imaged independently using the 0.55 and 1.56 THz configurations. The acquired images comprised \(2048 \times 2048\) pixels and spanned a field of view of \(50~\mathrm{mm} \times 50~\mathrm{mm}\) at the object plane, yielding a calibrated spatial sampling interval of \(24.41~\mu\mathrm{m/pixel}\). The intensity profiles extracted from the pinhole images were fitted with Lorentzian functions, with the results summarized in Fig. 4. At 0.55 THz, the measured FWHMs of the imaging spots are 1.432 mm along the x direction and 1.202 mm along the y direction. In contrast, the corresponding FWHMs at 1.56 THz decrease to 0.314 mm (x) and 0.354 mm (y), respectively. In the present setup, the THz imaging lens has a focal length f = 35 mm and an aperture D = 50.8 mm, giving an effective numerical aperture NA = 0.59. At 1.56 THz, the corresponding theoretical diffraction-limited FWHM is 0.20 mm, which serves as the reference for the measured spatial response. The measured values therefore reach approximately 1.6–1.8 times of this limit, consistent with the expected broadening from the finite 0.5 mm pinhole, laser-sheet non-uniformity, and atomic-medium effects.

To further validate the superior spatial resolution achieved at 1.56 THz compared with 0.55 THz, metallic masks featuring various Greek-letter patterns and a spiral target were imaged at both frequencies. As presented in Fig. 5, the 0.55 THz images exhibit pronounced spatial blurring and the coalescence of adjacent structures. In contrast, the 1.56 THz images faithfully preserve the sharp contours and narrow gaps of the Greek-letter patterns while resolving a significantly greater number of closely spaced turns in the spiral target. These observations align well with the preceding FWHM measurements, further confirming the enhanced recovery of high-spatial-frequency features at 1.56 THz. This performance gain stems from the shorter operating wavelength, which reduces the diffraction-limited spot size for a given numerical aperture—in full agreement with theoretical predictions of frequency-dependent spatial resolution.

\begin{figure}[htbp]
\centering\includegraphics[width=12cm]{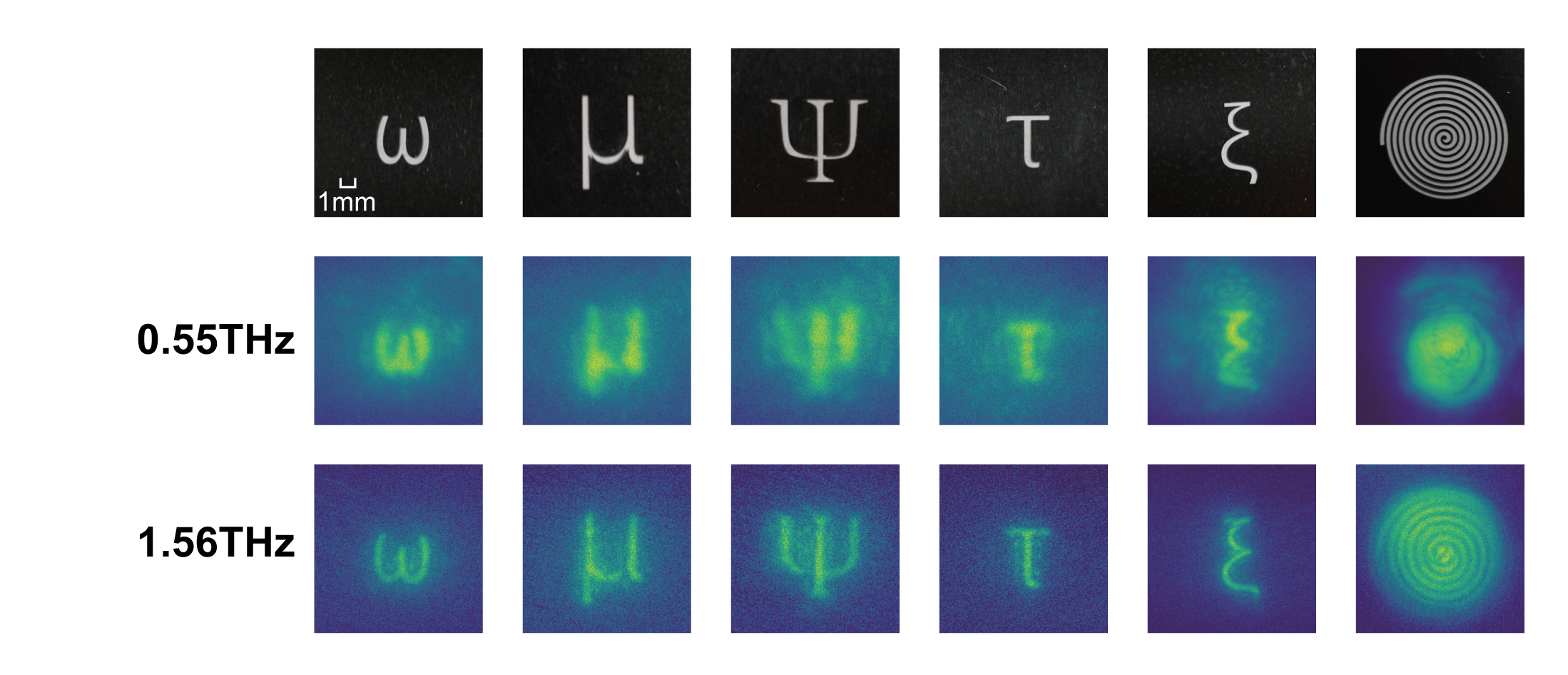}
\caption{Comparison of fluorescence-based THz imaging performance at 0.55 THz and 1.56 THz using metallic masks featuring various Greek-letter patterns and a spiral target. The top row displays photographs of the target masks, while the subsequent rows present the corresponding THz images acquired at 0.55 THz and 1.56 THz, respectively. }
\end{figure}

To evaluate the high-speed imaging performance of the 1.56 THz system, a rotating optical chopper (Thorlabs MC2000B) equipped with a 100-slot MC1F100 blade was employed as a dynamic target. The chopping frequency was set to 200 Hz, while fluorescence images were captured using a high-speed camera at 300 fps. As illustrated in Fig. 6, the position of the blade evolves distinctly across consecutive frames, enabling the direct visualization of its rotational motion. Although the current frame rate of 300 fps is restricted by the limited available power at 1.56 THz and the further reduced power density associated with a large field of view, these results successfully validate the high-speed dynamic imaging capability of the system. Such power constraints can be readily mitigated in future work by integrating our recently developed high-power THz sources \cite{bib23}.

\begin{figure}[htbp]
\centering\includegraphics[width=12cm]{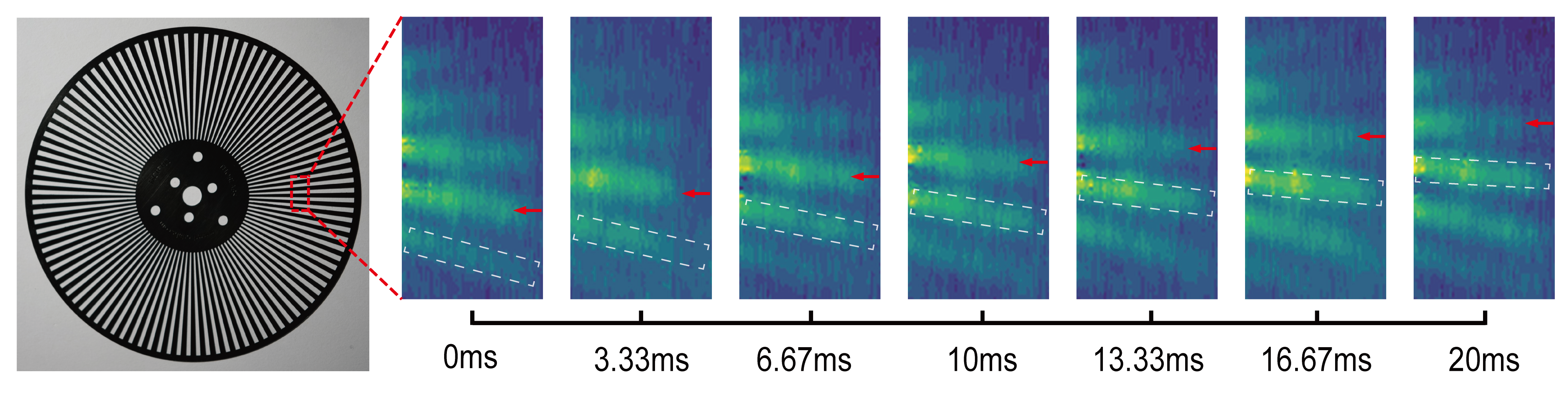}
\caption{High-speed imaging of a rotating optical chopper at 1.56 THz. An optical chopper equipped with a 100-slot blade was operated at a chopping frequency of 200 Hz, with fluorescence images captured at 300 fps. Representative consecutive frames clearly resolve the transient motion of the rotating blade, demonstrating the capability of the 1.56 THz imaging system for high-speed dynamic capture.}
\end{figure}

As shown in Fig.~7, the histopathological characteristics of the tumor tissue were investigated in conjunction with terahertz transmission imaging based on Rydberg-atom detection. Fig.~7(a) shows the terahertz transmission image acquired from a 200-$\mu$m-thick fresh tumor tissue section, which was experimentally determined to provide optimal terahertz imaging performance. Fig.~7(b) presents the corresponding H\&E-stained histological image obtained from a 30-$\mu$m-thick tissue section, as thinner sections facilitate dye penetration and improve the visualization of cellular details. Based on the terahertz transmission image in Fig.~7(a), image segmentation was performed using a threshold-based method, as shown in Fig.~7(c), where blue and white indicate smooth muscle/stromal and lymphocyte-rich regions, respectively. Furthermore, pixel-level tissue classification was performed on the H\&E-stained image in Fig.~7(b) using the Pixel Classification function in QuPath, as shown in Fig.~7(d). In this approach, representative tissue regions were manually annotated as training samples, enabling the classifier to learn image features, including color and texture, and subsequently assign individual pixels to different tissue classes. The green and purple regions represent smooth muscle/stromal and lymphocyte-rich regions, respectively. 

\begin{figure}[htbp]
\centering\includegraphics[width=7cm]{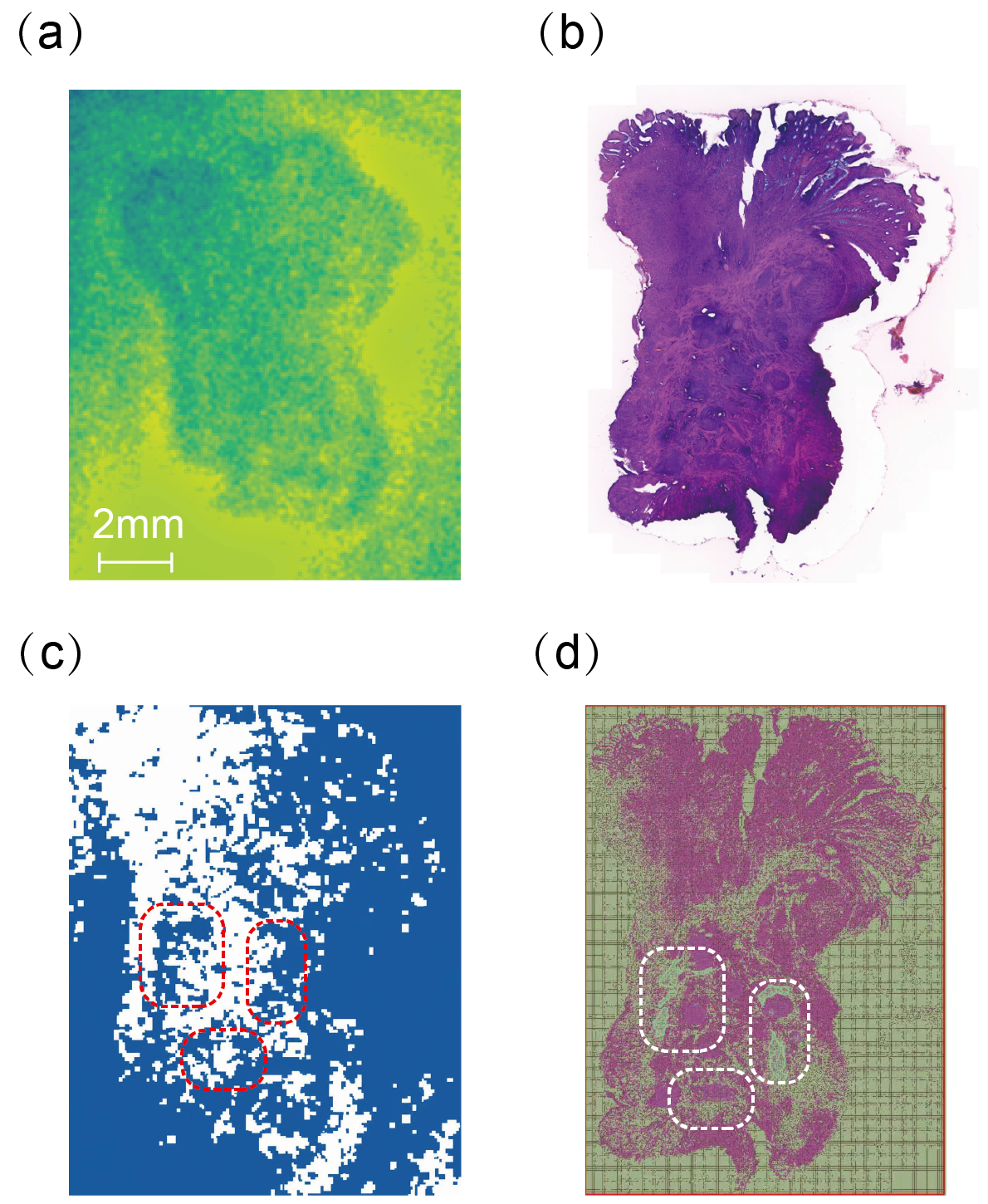} 
\caption{Histopathological analysis and Rydberg-atom-based terahertz transmission imaging of a fresh tumor tissue section.  
(a) THz transmission imaging of the tissue section acquired using Rydberg-atom-based detection, showing the spatial variations response to THz transmission. 
(b) Hematoxylin and eosin (H\&E)-stained image of the tissue section, showing its overall histopathological architecture and cellular distribution.
(c) Segmentation result of the THz image obtained using a threshold-based method, where blue and white represent smooth muscle/stromal and lymphocyte-rich regions, respectively.
(d) Pixel classification of the H\&E-stained image using the machine-learning-based Pixel Classification function in QuPath, illustrating the spatial distribution of different tissue components. Smooth muscle/stromal regions are labeled in green, whereas lymphocyte-rich regions are labeled in purple.}
\end{figure}

Comparison of Figs.~7(c) and 7(d) reveals similar morphological distributions of the identified tissue components, indicating a correspondence between the histopathological structures and the terahertz transmission patterns. Notably, the lymphocyte-rich regions exhibit lower terahertz transmission intensity than the smooth muscle/stromal regions. This difference may be attributed to the higher cellular density and distinct cellular architecture of lymphocyte-rich regions. Specifically, lymphocytes are characterized by a high nuclear-to-cytoplasmic ratio and relatively scant cytoplasm, resulting in densely packed nuclei and compact chromatin structures. These features may produce pronounced spatial variations in the effective refractive index, enhancing electromagnetic scattering and consequently reducing terahertz transmission. It is worth noting that the spatial positions of the corresponding regions in Figs.~7(c) and 7(d) are not perfectly aligned. This discrepancy may be attributed to the different thicknesses of the tissue sections used for histological staining (30~$\mu$m) and terahertz imaging (200~$\mu$m). During sectioning, the different thicknesses may result in distinct mechanical responses to shear forces, leading to varying degrees of tissue displacement. Consequently, regions with similar morphological features may exhibit positional shifts between the two images. Despite this spatial deviation, the morphological correspondence and distinct transmission contrast demonstrate the potential of Rydberg-atom-based THz imaging for differentiating tissue components. Unlike conventional histopathological analysis, which requires multiple tissue preparation and staining procedures, this technique offers rapid, label-free tissue characterization without the need for staining. These capabilities highlight its potential as a complementary tool for rapid pathological assessment, tissue composition mapping, and, ultimately, intraoperative tissue evaluation.

\section{DISCUSSION}
In summary, we have experimentally demonstrated a full-field THz imaging system operating at 1.56 THz by utilizing the \(11P_{3/2}\rightarrow10D_{5/2}\) transition of cesium Rydberg atoms, which provides a favorable balance among THz frequency, source power, excitation efficiency, THz coupling, and fluorescence yield. The system exhibits a response bandwidth of 24.37 MHz, a frame rate of 300 fps, and an exceptional sensitivity of $31.7~\mathrm{fW\,s^{-1/2}\,(6.5 \times 6.5~\mu m^{2})^{-1}}$ . After minor background subtraction and flat-field correction, the system achieves sub-millimeter spatial resolution, with measured spatial response FWHMs of 0.314 mm and 0.354 mm along the horizontal and vertical directions, respectively. These results approach 1.6–1.8 times of the theoretical diffraction-limited resolution, demonstrating the substantial improvement in spatial resolving capability enabled by high-frequency operation.
These findings demonstrate that scaling to higher frequencies provides a direct and effective pathway to enhancing the far-field spatial resolution of Rydberg-atom-based THz imaging while preserving single-shot, full-field optical readout. Specifically, targets imaged at 1.56 THz exhibit sharper boundaries and clearer feature separation than those at 0.55 THz, confirming that the shorter operating wavelength is successfully translated into sub-millimeter spatial resolution without mechanical raster scanning.

Looking forward, the performance of the current 1.56 THz imaging system is primarily constrained by the characteristics of the available THz source. In particular, the limited THz power restricts the strength of the THz-induced fluorescence response, thereby requiring relatively long camera exposure times to maintain an adequate signal-to-noise ratio. Consequently, although the present system demonstrates full-field image acquisition at 300 fps, its acquisition speed is not fundamentally limited by the intrinsic atomic response. Integrating higher-power THz sources is expected to enhance the fluorescence signal and enable shorter exposure times, thereby providing a direct route toward higher imaging frame rates. Nevertheless, the ultimate acquisition speed will also depend on fluorescence collection efficiency, atomic motion, camera readout noise, and data-transfer bandwidth. Furthermore, extending the operating frequency to 5 THz could push the spatial resolution to the tens-of-micrometers scale. Advanced large-scale facilities, such as free-electron lasers, may further extend the frequency range to 7–30 THz, offering new opportunities for micrometer-scale THz imaging \cite{bib23,bib24}.

Moreover, the distinct spectral selectivity observed across different Rydberg transitions opens up promising avenues for multi-color THz imaging, where multiple THz frequencies can be discriminated via unique fluorescence signatures and selectively isolated using appropriate optical bandpass filters. Finally, while our rotating-chopper measurements confirm rapid full-field acquisition, and the intrinsic lifetime of the Rydberg state permits an intrinsic response bandwidth approaching the megahertz regime, the practical frame rate remains bounded by atomic motion, fluorescence collection efficiency, camera exposure time, readout noise, and data bandwidth. Overcoming these limitations will pave the way for ultrahigh-speed, real-time THz imaging applications.

These preliminary tissue-slice measurements highlight the promise of our enhanced spatial resolution for biological imaging. The observed THz contrast likely stems from spatial variations in water content, density, dielectric response, and tissue composition. However, the present results should be interpreted as an initial demonstration of structural contrast rather than definitive evidence of diagnostic specificity. Correlative comparisons with histological sections, measurements across larger sample cohorts, and quantitative analyses of normal versus pathological regions will be essential to decouple disease-specific signatures from physiological variations in sample thickness, hydration, and morphology. Ultimately, this robust and rapid full-field technique with high spatial resolution lays the groundwork for future
real-time biomedical diagnostics, non-destructive evaluation, and dynamic THz wave visualization.

\section*{Acknowledgment}
This work was supported by the CAS Project for Young Scientists in Basic Research (YSBR-042), the National Natural Science Foundation of China (12125508, 12541503), and the International Cooperation Program, Shanghai Municipal Commission of Science and Technology (26140780400), the China Postdoctoral Science Foundation under Grant No.~GZC20250217, the Basic Research Program of Science and Technology Commission of Shanghai Municipality (25ZR1402527).

\bibliography{citation}



\end{document}